\documentclass[prd,superscriptaddress,altaffilletter,nofootinbib]{revtex4-1}

\usepackage{hyperref}
\usepackage{amsmath}
\usepackage{amssymb}
\usepackage{amsthm}
\usepackage{amsfonts}
\usepackage[margin=1.0in]{geometry}
\usepackage[light]{kpfonts}
\usepackage{listings}
\usepackage{microtype}
\usepackage{siunitx}
\usepackage[]{graphicx}
\usepackage{caption}
\usepackage{subcaption}

\theoremstyle{remark}

\usepackage{comment}
\usepackage{hyperref}
\usepackage[nolist]{acronym}
\usepackage[usenames,dvipsnames]{xcolor}
\usepackage{xspace}
\usepackage{tikz}
\usepackage{textcomp}
\usepackage[export]{adjustbox}
\usetikzlibrary{arrows,shapes,trees,decorations.pathreplacing}
\allowdisplaybreaks

\begin{document}

\title{A self-consistent method to estimate the rate of compact binary coalescences with a Poisson mixture model}

\author{Shasvath J. Kapadia}
\affiliation{Leonard E.\ Parker Center for Gravitation, Cosmology, and Astrophysics, University of Wisconsin-Milwaukee, Milwaukee, WI 53201, USA}

\author{Sarah Caudill}
\affiliation{Nikhef, Science Park, 1098 XG Amsterdam, Netherlands}

\author{Jolien D. E. Creighton}
\affiliation{Leonard E.\ Parker Center for Gravitation, Cosmology, and Astrophysics, University of Wisconsin-Milwaukee, Milwaukee, WI 53201, USA}

\author{Will M. Farr}
\affiliation{Department of Physics and Astronomy, Stony Brook University, Stony Brook NY 11794, USA}

\author{Gregory Mendell}
\affiliation{LIGO Hanford Observatory, Richland, WA 99352, USA}

\author{Alan Weinstein}
\affiliation{LIGO Laboratory, California Institute of Technology, MS 100-36, Pasadena, California 91125, USA}

\author{Kipp Cannon}
\affiliation{Canadian Institute for Theoretical Astrophysics, 60 St. George Street, University of Toronto, Toronto, Ontario, M5S 3H8, Canada}
\affiliation{RESCEU, The University of Tokyo, Tokyo, 113-0033, Japan}

\author{Heather Fong}
\affiliation{Canadian Institute for Theoretical Astrophysics, 60 St. George Street, University of Toronto, Toronto, Ontario, M5S 3H8, Canada}
\affiliation{Department of Physics, 60 St. George Street, University of Toronto, Toronto, Ontario, M5S 3H8, Canada}
\affiliation{RESCEU, The University of Tokyo, Tokyo, 113-0033, Japan}

\author{Patrick Godwin}
\affiliation{Department of Physics, The Pennsylvania State University, University Park, PA 16802, USA}
\affiliation{Institute for Gravitation and the Cosmos, The Pennsylvania State University, University Park, PA 16802, USA}

\author{Rico K. L. Lo}
\affiliation{LIGO Laboratory, California Institute of Technology, MS 100-36, Pasadena, California 91125, USA}
\affiliation{Department of Physics, The Chinese University of Hong Kong, Shatin, New Territories, Hong Kong}

\author{Ryan Magee}
\affiliation{Department of Physics, The Pennsylvania State University, University Park, PA 16802, USA}
\affiliation{Institute for Gravitation and the Cosmos, The Pennsylvania State University, University Park, PA 16802, USA}

\author{Duncan Meacher}
\affiliation{Leonard E.\ Parker Center for Gravitation, Cosmology, and Astrophysics, University of Wisconsin-Milwaukee, Milwaukee, WI 53201, USA}

\author{Cody Messick}
\affiliation{Department of Physics, The Pennsylvania State University, University Park, PA 16802, USA}
\affiliation{Institute for Gravitation and the Cosmos, The Pennsylvania State University, University Park, PA 16802, USA}

\author{Siddharth R. Mohite}
\affiliation{Leonard E.\ Parker Center for Gravitation, Cosmology, and Astrophysics, University of Wisconsin-Milwaukee, Milwaukee, WI 53201, USA}
\affiliation{LSSTC Data Science Fellow}

\author{Debnandini Mukherjee}
\affiliation{Leonard E.\ Parker Center for Gravitation, Cosmology, and Astrophysics, University of Wisconsin-Milwaukee, Milwaukee, WI 53201, USA}

\author{Surabhi Sachdev}
\affiliation{Department of Physics, The Pennsylvania State University, University Park, PA 16802, USA}
\affiliation{Institute for Gravitation and the Cosmos, The Pennsylvania State University, University Park, PA 16802, USA}

\begin{abstract}
The recently published GWTC-1 \cite{GWTC-1} - a journal article summarizing the search for gravitational waves (GWs) from coalescing compact binaries in data produced by the LIGO-Virgo network of ground-based detectors during their first and second observing runs -  quoted estimates for the rates of binary neutron star, neutron star black hole binary, and binary black hole mergers, as well as assigned probabilities of astrophysical origin for various significant and marginal GW candidate events. In this paper, we delineate the formalism used to compute these rates and probabilities, which assumes that triggers above a low ranking statistic threshold, whether of terrestrial or astrophysical origin, occur as independent Poisson processes. In particular, we include an arbitrary number of astrophysical categories by redistributing, via mass-based template weighting, the foreground probabilities of candidate events, across source classes. We evaluate this formalism on synthetic GW data, and demonstrate that this method works well for the kind of GW signals observed during the first and second observing runs. 
\end{abstract}

\maketitle

\section{Introduction}\label{sec:intro}

The detection of gravitational waves from a coalescing black hole binary, on September 14, 2015, by the Advanced LIGO (Laser Interferometer Gravitational-Wave Observatory) network of ground based detectors, announced the arrival of gravitational-wave astronomy \cite{GW150914-DETECTION, GW150914-CBC}. Since then, a number of additional gravitational-wave detections have been made. These include 10 from merging binary black holes (BBHs) \cite{GW151226-DETECTION, GW170104-DETECTION, GW170608-DETECTION, GW170814-DETECTION, GWTC-1} and 1 from a coalescing binary neutron star \cite{GW170817-DETECTION} that also produced an electromagnetic counterpart amply identified by a number of telescopes worldwide \cite{GW170817-MMA}. All of these detections are reported in ``GWTC-1: A Gravitational-Wave Transient Catalog of Compact Binary Mergers Observed by LIGO and Virgo during the First and Second Observing Runs'' \cite{GWTC-1}. We will refer to these observing runs as O1 and O2.

A plethora of interesting astrophysical questions can now be asked, and answered with appropriately quantified uncertainties. Among them is the question: how many compact binary coalescence (CBC) events, of a specific astrophysical source class, occur per unit spacetime volume? The determination of these coalescence rates is a high-profile science target of the LIGO-Virgo collaboration, and for good reason. Prior to the first gravitational-wave detection, estimates of BBH merger-rates spanned many orders of magnitude \cite{Rates-pre-detection}. These were based on population models that remained unconstrained due to a paucity of electromagnetic observations. GW events allow these rates to be better constrained in a largely model-independent manner. Using the results from O1 and O2, credible intervals and upper limits on the rate of binary black hole (BBH), neutron star - black hole (NSBH) binary, and binary neutron star (BNS) mergers have been previously published in \cite{GW150914-RATES, GW150914-RATES-SUPPLEMENT, O1:BBH, O1-RATES-UL, GW170104-DETECTION, GW170817-DETECTION}.

The rates (at $90\%$ confidence) for the same astrophysical source classes were re-estimated after O2,  details of which may be found in \cite{GWTC-1}, which we shall hereafter call the ``catalog''. The BNS merger rate was updated to $110-3840~\mathrm{Gpc}^{-3}\mathrm{yr}^{-1}$, the BBH merger rate to $ 9.7-101~\mathrm{Gpc}^{-3}\mathrm{yr}^{-1}$, and a rate upper limit for NSBH mergers was placed at $610~\mathrm{Gpc}^{-3}\mathrm{yr}^{-1}$. The goal of this write-up is to serve as a complement to the catalog. In particular, we delineate the formalism and method used to determine the rates of compact binary mergers, and probabilities of astrophysical origin for various highly significant events in O1 and O2, quoted in the catalog for the GstLAL search \cite{GSTLAL-1}.

The method we present in this paper is a non-trivial extension of one that was developed in the context of gravitational wave data analysis by W. Farr, J. Gair, I. Mandel, and C. Cutler (henceforth FGMC) \cite{FGMC}. (Note that a related formalism was developed earlier, in the context of gamma-ray burst data analysis \cite{Loredo-Wasserman}). The FGMC formalism itself is an application of Poisson statistics to determine the rate of astrophysical events when supplied with a mixture of foreground and background events, given that the foreground and background models are known (or assumed) even though the membership of each event to either one of these classes is unknown. 

The FGMC formalism was employed in the determination of BBH rates using data from O1 \cite{GW150914-RATES}. The scarcity of confirmed BBH detections prompted the use of a method prescribed by Kim et al \cite{Kim-et-al} where each BBH detected was assumed to be a distinct astrophysical source. On the other hand, the rate upper limits of BNS and NSBH mergers were computed by formally assuming that exactly zero BNS and NSBH events occurred during observation time, and then employing the Poisson distribution for zero events \cite{O1-RATES-UL}. This assumption is based on the fact that no BNS and NSBH events, with a false alarm rate (FAR) of less than 1/100 years, were found. While changing the threshold by a few orders of magnitude does not vastly alter the results \cite{O1-RATES-UL}, the choice of threshold is in itself somewhat arbitrary. 

What we propose, as an alternative to what was done in O1, is a self consistent, threshold-independent, counting method that simultaneously estimates the rates of BNS, NSBH and BBH mergers. In essence, this method extends the FGMC formalism by constructing a joint posterior on the Poisson expected counts for an arbitrary choice of foreground categories by redistributing, via mass-based template-weighting, the foreground probabilities of candidate events across astrophysical source classes. Thus, the method presented here, while used in the catalog paper to handle the three astrophysical categories mentioned above, can handle any number of categories. For example, it could set limits on the rates of BBH mergers in the proposed mass gaps \cite{O2:RP} and possibly including a higher mass black hole region too.

The template-weights themselves are computed from simulation runs (software injection campaigns), each targeted at a distinct astrophysical source type. An injection campaign involves adding synthetic signals pertaining to a source class with clearly defined mass and spin distributions, into the detector data, and recovering them via a detection pipeline. To construct the weights, we count how many times injections of a given category are recovered in a given template bin and divide this by the
total number of recovered injections pertaining to that category.

In the following sections, we describe this ``multi-component'' extension of the FGMC method, by first constructing the joint posterior on the Poisson expected counts for each source category, followed by details on how to estimate the spacetime volume sensitivities of the detectors to each of the source categories and how to incorporate uncertainties in their measurements into the rates posterior. We then apply this extension to synthetic data, in order to assess its ability to accurately categorize and count a mixture of CBC signals. We end by summarizing the multi-component FGMC method, evaluating its performance, and suggesting other applications of the method.

\section{Constructing the multi-component counts posterior from candidate events}\label{sec:posterior}

\subsection{Posterior on Terrestrial and Astrophysical counts}
The original FGMC method \cite{FGMC} constructs a two-component likelihood on the expected number of astrophysical ($\Lambda_1$) and terrestrial ($\Lambda_0$) counts, per experiment, assuming that the foreground and background triggers above a low ranking statistic threshold where background triggers \footnote{A trigger is a gravitational wave candidate event acquired during a templated matched-filtering based analysis of detector strain data. A background trigger is one that was most likely produced by terrestrial processes.} dominate, occur as independent Poisson processes. More specifically, the expected counts ($\Lambda$) referred to here are the Poisson means for the duration of the experiment (the total observing time), given which one can compute the discrete probability distribution on the number $k$ of occurrences of events :
\begin{equation}
p(k | \Lambda) \propto \Lambda^k\exp(-\Lambda).
\end{equation}
The two-component FGMC likelihood has the following form \cite{FGMC}:
\begin{equation}
p(\vec{x} | \Lambda_0, \Lambda_1) = e^{-\Lambda_0-\Lambda_1}\prod_{j=1}^{N}[\Lambda_0 b(x_j) + \Lambda_1 f(x_j)]
\end{equation}
where $\vec{x} = \lbrace x_j \rbrace, j = 1, 2, 3, ..., N$, are the ranking statistics of the triggers above threshold, and $b(x_j), f(x_j)$ are the background and foreground distributions (normalized density functions, also called the background and foreground models), evaluated at $x_j$ ($b(x_j) \equiv p(x_j | \mathrm{noise})$ and $f(x_j) \equiv p(x_j | \mathrm{signal})$). It is worth noting here that the foreground count $\Lambda_1$ is directly proportional to the astrophysical rate of mergers $R$, which can be determined if the population averaged spacetime volume sensitivity $\langle VT \rangle$ of the detector is known:
\begin{equation}
\Lambda_1 = R\cdot \langle VT \rangle.
\end{equation}
The FGMC likelihood may therefore also be written in terms of $R$ and $\langle VT \rangle$. One is thus at liberty to choose a prior, either on $\Lambda_1$, or on $R$ and $VT$ jointly. We will come back to this in an upcoming section where we discuss incorporating uncertainties in the measurement of the spacetime volume sensitivity into the rates posteriors. For the moment, we proceed by choosing a prior on $\Lambda_0, \Lambda_1$, and writing the FGMC posterior \cite{FGMC, GW150914-RATES}:
\begin{equation}
p(\Lambda_0, \Lambda_1 | \vec{x}) \propto p(\Lambda_0, \Lambda_1)e^{-\Lambda_0-\Lambda_1}\prod_{j=1}^{N}[\Lambda_0 b(x_j) + \Lambda_1 f(x_j)]
\end{equation}
where $p(\Lambda_0, \Lambda_1)$ is the prior on the counts. We wish to extend this method to include an arbitrary number of astrophysical components (BNS, NSBH, BBH, possibly others), in place of a single aggregated astrophysical component.

\subsection{Multi-component counts posterior}

To a very good approximation, the foreground distribution of ranking statistics is independent of source category \cite{GW150914-RATES-SUPPLEMENT, Schutz2011, ChenAndHolz}. Symbolically:
\begin{equation}
p(x | \alpha) \approx p(x | \mathrm{signal}) 
\end{equation}
where $\alpha$ is an astrophysical source category. However, it is necessary to re-weight the foreground distribution with source-category specific weights $W_{\alpha}(x)$. This re-weighting would allow us to split the foreground distribution into multiple, source-specific foreground distributions $f_{\alpha}(x)$, where $x$ could now encapsulate multiple properties of a trigger, and not necessarily only the ranking statistic. The general mathematical form of the posterior becomes:
\begin{equation}
p(\Lambda_0, \vec{\Lambda}_1 | \vec{x}) \propto p(\Lambda_0, \vec{\Lambda}_1)e^{-\Lambda_0-\vec{\Lambda}_1\cdot\vec{u}}\cdot\prod_{j=1}^{N}[\Lambda_0 b(x_j) + \vec{f}(x_j)\cdot\vec{\Lambda}_1]
\end{equation}
where $\vec{f}(x) \equiv \lbrace f_{\alpha}(x) \rbrace ~( \mathrm{for}~\alpha = \lbrace{\mathrm{BNS, NSBH}, \ldots \rbrace})$ is a vector of source-specific foreground distributions, $\vec{\Lambda}_1 \equiv \lbrace \Lambda_{\alpha}\rbrace ~(\mathrm{for}~\alpha = \lbrace{\mathrm{BNS, NSBH},\ldots \rbrace)}$ is a vector of source-specific Poisson expected counts, and $\vec{u}$ is a vector of 1s (See Appendix~\ref{sec:appA}).

Gravitational waves from different CBC sources are expected to activate (ring-up) templates from different (though not necessarily disjoint) regions of a template-bank's parameter space. By dividing the template bank into multiple bins (which we denote as $m$), we can assign to each trigger source-specific template-weights based on the bin in which the template lives. As derived in Appendix \ref{sec:appA}, the source-specific foreground distributions are constructed using template-weights and bin-dependent foreground distributions:
\begin{equation}
f_{\alpha}(x) \equiv p(L, m~|~\alpha) \approx p(L~|~m,\mathrm{signal})\cdot W_{\alpha}(m).
\end{equation}
Assuming that the detector data was analyzed with the GstLAL detection pipeline \cite{GSTLAL-1, GSTLAL-2}, $L$ is the log-likelihood-ratio ranking statistic \cite{GSTLAL-LR}, $m$ is the bin number, $W_{\alpha}(m) \equiv p(m~|~\alpha)$ are the bin-dependent template weights, and $p(L~|~m,\mathrm{signal})$ are the bin-dependent foreground distributions. On the other hand, the background distribution is given by:
\begin{equation}
b(x) \equiv p(L,m~|~\mathrm{noise}) = p(L~|~m,\mathrm{noise})\cdot W_0(m)
\end{equation}
where $W_0(m) \equiv p(m~|~\mathrm{noise})$.

It is convenient to define source-specific Bayes-factors for a trigger $x$, using the foreground and background distributions evaluated at $x$'s ranking statistic value $L$, as well as the template weights:
\begin{equation}
\vec{K}(x) \equiv \frac{\vec{f}(x)}{b(x)}=\frac{p(L~| m, \mathrm{signal})}{p(L~|~m, \mathrm{noise})}\frac{\vec{W}_1(m)}{W_0(m)}
\end{equation}
where $\vec{W}_1(m) = \lbrace W_{\mathrm{BNS}}(m), W_{\mathrm{NSBH}}(m), \ldots \rbrace$. The multi-component counts posterior can now be written more compactly as:
\begin{equation}\label{counts_posterior}
p(\Lambda_0, \vec{\Lambda}_1 | \vec{x}) \propto p(\Lambda_0, \vec{\Lambda}_1)e^{-\Lambda_0-\vec{\Lambda}_1\cdot \vec{u}}\cdot\prod_{j=1}^{N}[\Lambda_0 +  \vec{\Lambda}_1\cdot \vec{K}(x_j)].
\end{equation}
\subsection{Useful Approximations to the Multi-component Counts Posterior}

In this section, we recast the counts posterior (\eqref{counts_posterior}) in approximate forms that make it computationally efficient to marginalize out the terrestrial count $\Lambda_0$. 

If the number of candidate events is sufficiently large ($N  >> 1$), and the number of background events vastly exceeds the number of foreground events, then, using the method of Laplace: 
\begin{equation}
\Lambda_0^N e^{-\Lambda_0}\approx N^Ne^{-N}e^{-\left(\Lambda_0-N \right)^2/(2N)}
\end{equation}
The multi-component counts posterior then assumes the form:
\begin{equation}\label{counts_posterior_cl}
p(\Lambda_0, \vec{\Lambda}_1 | \vec{x}) \propto N^Ne^{-N}e^{-\left(\Lambda_0-N \right)^2/(2N)}p(\Lambda_0, \vec{\Lambda}_1)e^{-\vec{\Lambda}_1\cdot \vec{u}}\cdot\prod_{j=1}^{N}[1 +  \frac{\vec{\Lambda}_1}{\Lambda_0}\cdot \vec{K}(x_j)] 
\end{equation}
From the perspective of determining astrophysical rates, the posterior on the terrestrial counts is generally not of much interest, and is usually marginalized out. Writing the multi-component posterior in the above form makes the marginalization over terrestrial counts amenable to Gauss-Hermite quadrature. 

A further simplification to the counts posterior can be written down, again in the limit of large $N$ dominated by background events. In this simplification, the posterior on the terrestrial count is modeled as a Dirac-delta function centered on $N$. Thus, when marginalizing out the terrestrial count, the multi-component counts posterior assumes the form:
\begin{equation}\label{counts_posterior_df}
p(\vec{\Lambda}_1 | \vec{x}) \propto p(N, \vec{\Lambda}_1)e^{-\vec{\Lambda_1}\cdot \vec{u}}\cdot\prod_{j=1}^{N}[1 + \vec{\Lambda}_1\cdot \vec{k}(x_j)]
\end{equation}
where $\vec{k}(x) \equiv \frac{\vec{K}(x)}{N}$ is the reduced Bayes factor for trigger $x$.

Note that $k_\alpha(x_j) << 1$ for the majority of triggers, since the majority are background events. Conversely, $k_\alpha(x_j) >> 1$ for certain highly significant foreground events, i.e, events that are almost unambiguously of astrophysical category $\alpha$. 

\subsection{Bin-dependent template weights}\label{template_weights}

The key to constructing the multi-component counts posterior is to determine the weights $W_{\alpha}(m) \equiv p(m | \alpha)$, which is a measure of how the astrophysical signals of a specific source category distribute themselves in the template bank. This subsection outlines how these weights are approximately computed for the GstLAL detection pipeline.

The GstLAL pipeline splits the template bank into sub-banks (which we simply refer to as bins) \cite{GSTLAL-SVD}, in the ``$\mathcal{M}-\chi_{\mathrm{eff}}$" space, where $\mathcal{M}$ is the template's chirp mass,  and $\chi_{\mathrm{eff}}$ is the template's effective spin parameter. The chirp mass is defined as:
\begin{equation}
\mathcal{M} = \frac{(m_1m_2)^{3/5}}{(m_1 + m_2)^{1/5}}
\end{equation}
and the effective spin parameter is defined as:
\begin{equation}
\chi_{\mathrm{eff}} = \frac{m_1\chi_1 + m_2\chi_2}{m_1 + m_2}
\end{equation}
with $m_1, m_2, \chi_1, \chi_2$ as the component masses and spin angular momenta (or more precisely, their components parallel to the orbital angular momentum) of the binary.
 
While this binning was originally designed to speed-up the extraction of GW signals from detector data, one can also exploit it for the construction of template-weights and the multi-component counts posterior.

These bins can be thought of as coarse-grained templates; when a template is ``rung-up'', the corresponding bin in which it lives is said to be ``activated''. Thus, during a run, we can count the number of times each bin gets activated, and thus determine an ``activation count'' for each bin. 
Intuitively, one can see that signals from a specific astrophysical source-class will tend to predominantly activate only a subset of all the bins. BNS signals for example will tend to activate low-mass bins, whereas BBH signals will tend to activate high-mass bins.  

Now, suppose we run distinct injection campaigns targeted at specific source categories (BNS, NSBH, BBH, ...). In other words, we inject simulated waveforms of a specific source class, and recover these injections using the binned template bank. Considering only those injections that were recovered with false alarm rate (FAR) of less than 1/30 days, we determine activation counts $A_{\alpha}, (\alpha = 1,2,...,Q,$ where $Q$ is the total number of astrophysical categories) corresponding to these injections, and construct a set of weights as follows:
\begin{equation}\label{weights}
W_\alpha (m) = \frac{A_{\alpha}(m)}{\sum_{m=0}^{N_{\mathrm{bins}}-1} A_{\alpha}(m)}
\end{equation}
where $m$ is the bin number and $N_{\mathrm{bins}}$ is the total number of bins. (Note that GstLAL-based CBC rate estimations quoted in the catalog paper were computed with analyses that split the template bank into 686 bins. A visual representation of these weights is shown in Figure~\ref{fig:activation_counts} .)

To define specific source categories, injection campaigns were designed
to reflect the choices made for parameter boundaries of the
astrophysical sources in the catalog paper. The BNS source category included neutron stars with component masses $m_i$ distributed uniformly in $\ln m_i$ between $1\le m_i/M_\odot \le 3.0$ such that the total mass $M$ was less than $6.0M_\odot$. The lower mass limit is motivated by a 6$\sigma$ deviation from masses of components in double neutron star systems~\cite{Ozel2016} while the upper mass limit is motivated by certain models and observations which allow neutron stars to form up to 3$M_\odot$~\cite{Rhoades1974, Kalogera1996, OzelNSLower, Lattimer2012, Kiziltan2013}. Spin vectors for BNS components were allowed to be isotropic in direction and uniform in magnitude, with a maximum allowed spin magnitude of 0.4. This maximum magnitude constraint is motivated by observations of the fastest spinning pulsar with $\chi \lesssim 0.4$~\cite{Hessels2006}. The BBH source category included black holes with component masses distributed uniformly in $\ln m_i$ between $5\le m_i/M_\odot \le 50$ such that $M\le 100M_\odot$. The lower mass limit is motivated by the possible existence of a minimum black hole mass~\cite{Bailyn1998, Ozel2010, Farr2011} while the upper mass limit is motivated by evidence of an upper cutoff in the BBH mass spectrum based on the first few LIGO detections~\cite{Fishbach2017, Talbot2018, Wysocki2018}. Spin vectors for BBH components were allowed to be isotropic on the sphere with a maximum allowed spin magnitude of 0.99. The relativistic Kerr bound provides a theoretical constraint on black hole spin magnitude of 1.0 although we are also constrained by the limit of the waveform approximant. The NSBH source category included neutron stars with component masses distributed uniformly in $\ln m_1$ between $1.0\le m_1/M_\odot \le 3.0$ and black holes with component masses distributed uniformly in $\ln m_2$ between $5.0\le m_2/M_\odot \le 100.0$. The total mass for the NSBH category was constrained to $M\le 103.0M_\odot$. Both the NS and BH components were allowed to be isotropic on the sphere with maximum allowed spin magnitudes of 0.4 and 0.99, respectively. These mass and spin limits are motivated by the NS and BH observations and theoretical constraints already mentioned.\footnote{To ensure appropriate coverage of the component mass space, an additional injection set where at least one of the components lies in the range $3-5 M_{\odot}$, was constructed. The masses were distributed uniformly in $\mathrm{ln}~m_i$, with the other component spanning $1-100 M_{\odot}$. The spins were assumed to be isotropic, with a maximum value of $0.4$ for the first component, and $0.99$ for the latter.}

Injections in each source category were distributed uniformly in co-moving volume out to redshift of 0.2 for BNS and NSBH and out to 0.7 for BBH. In order to maximize the number of recoverable injections included in the injection campaign, an initial cut on expected signal-to-noise ratio less than 3.0 was applied to exclude injections that would be too far away or in a poor sky location for either of the Hanford or Livingston detectors. The parameters of these injections were tabulated and stored as unrecoverable.

\subsection{Probability of Astrophysical Origin}\label{sec:pastro}

From the original, two-component, FGMC counts posterior, one can compute the posterior probability that an event, with foreground and background distribution values $f(x)$ and $b(x)$, evaluated at the event's ranking statistic $x$, is of astrophysical origin, given the data $\vec{x}$ \cite{GW150914-RATES}:
\begin{equation}
P_1(x~|~\vec{x}) = \int_0^{\infty}p(\Lambda_0,\Lambda_1~|~\vec{x})\frac{\Lambda_1f(x)}{\Lambda_0b(x) + \Lambda_1f(x)}d\Lambda_0d\Lambda_1.
\end{equation}
Its complementary quantity is the posterior probability $P_0(x~|~\vec{x})$ that the same event originated from the Earth, with $P_0(x) + P_1(x) = 1$. 

These posterior probabilities can be straightforwardly extended to the case when we have a multi-component counts posterior. The source-specific foreground distributions of the multi-component posterior allow one to compute posterior probabilities pertaining to specific astrophysical source categories: 
\begin{equation}\label{p_category}
P_{\alpha}(x | \vec{x}) = \int_{0}^{\infty}p(\Lambda_0,\vec{\Lambda}_1 | \vec{x})\frac{\Lambda_{\alpha}K_{\alpha}(x)}{\Lambda_0 + \vec{\Lambda}_1\cdot\vec{K}(x)}d\Lambda_0 d\vec{\Lambda}_1.
\end{equation}
The complementary terrestrial posterior probability is once again related to the astrophysical probabilities via $P_0(x|\vec{x}) + \sum_{\alpha}P_{\alpha}(x~|~\vec{x}) = 1$, where $\alpha$ is summed over all astrophysical source categories corresponding to the multi-component posterior.

Astrophysical probabilities of candidate events are of considerable interest, from the perspective of following up gravitational wave events with telescopes sensitive to various parts of the electromagnetic spectrum. For instance, if $P_{\mathrm{BNS}}$ were high, the probability that this event would produce an electromagnetic counterpart also becomes high; such information could therefore be of great value to astronomers interested in hunting for electromagnetic signals associated with GWs, if reported with a sufficiently low latency from the time of occurrence of a GW candidate event.

With low-latency in mind, it is possible to re-write astrophysical probabilities as a function of the mean values of the Poisson expected counts. We define the mean values in the standard way:
\begin{equation}
\langle \Lambda_{\alpha} \rangle = \int_0^{\infty}\Lambda_{\alpha}p(\Lambda_0,\vec{\Lambda}_1|~\vec{x})d\Lambda_0d\vec{\Lambda}_1
\end{equation}
Suppose now we have a set of N candidate events $\vec{x}_N = \lbrace x_0, x_1,\ldots,x_{N-1}\rbrace$, from which we compute the mean value for the terrestrial Poisson count $\langle \Lambda_0 \rangle_N$ and astrophysical Poisson counts $\langle \vec{\Lambda}_1 \rangle_N = \lbrace {\langle\Lambda_1\rangle, \langle \Lambda_2 \rangle, \ldots, \langle \Lambda_Q \rangle \rbrace}$. The astrophysical probability, for category $\alpha$, of a new candidate event $x_{N+1}$ can now be computed as (See Appendix \ref{sec:appB}):
\begin{equation}
P_{\alpha}(x_{N+1}|\vec{x}_{N+1}) = \frac{\langle \Lambda_{\alpha} \rangle_N K_{\alpha}(x_{N+1})}{\langle \Lambda_0 \rangle_N + \langle\vec{\Lambda}_1\rangle_N\cdot \vec{K}(x_{N+1})}
\end{equation}
The above expression can be readily derived using Eq.~\eqref{p_category} and Eq.~\eqref{updated_posterior}.
Thus, if the mean values $\langle \Lambda_0 \rangle_N$ and $\langle \vec{\Lambda}_1\rangle_N$ are pre-computed, then the the astrophysical probabilities $P_{\alpha}(x_{N+1}|\vec{x}_{N+1})$ for a new candidate event can be computed almost instantaneously using only a handful of floating point operations. (Of course, the mean values would then need to be updated using Eq.~\eqref{updated_mean}).

\section{Determining the spacetime-volume sensitivity $\langle VT \rangle$}\label{sec:vt_sensitivity}

In order convert the posterior on counts to a posterior on rates, we need to determine the population averaged spacetime volume sensitivity $\langle VT \rangle_{\alpha}$ of the detectors, for every astrophysical source category $\alpha$. This sensitivity is written as \cite{GWTC-1}:
\begin{equation}\label{VT_integral}
\langle VT \rangle_{\alpha} = T\int dzd\theta \frac{dV_c}{dz}\frac{1}{1+z}p_{\alpha}(\theta)f(z,\theta)
\end{equation}
where T is the duration over which the sensitivity is estimated, z is the redshift, $V_c$ is the co-moving volume, $p_{\alpha}(\theta)$ is an assumed distribution of source-parameters $\theta$ for source-category $\alpha$, and $f(z,\theta)$ is a selection function that estimates how likely it is to detect sources with parameters $\theta$ at redshift $z$.

The above quantity is typically estimated using Monte-Carlo integration. Simulated signals (injections), with parameters drawn from $p_{\alpha}(\theta)$ and placed in redshift assuming standard cosmology, are added to the detector data. They are then searched for by the GstLAL pipeline, and assigned Bayes factors $\vec{K}(x)$. An estimate is then made of the number of injections that were recovered, $N_{\mathrm{rec}}$. Since the number that was injected, $N^{\mathrm{inj}}_{\alpha}$, and the spacetime volume into which the injections were made, $\langle VT \rangle^{\mathrm{inj}}_{\alpha}$, are both known, the measured spacetime volume is simply the injected spacetime volume re-scaled by the ratio of number-recovered to number-injected \cite{Tiwari2018}:
\begin{equation}
\langle VT \rangle_{\alpha} = \frac{N_{\mathrm{rec}}}{N^{\mathrm{inj}}_{\alpha}}\langle VT \rangle^{\mathrm{inj}}_{\alpha}.
\end{equation}
The crucial step then in the evaluation of the sensitive spacetime volume is the determination of $N_{\mathrm{rec}}$. In order to be consistent with the way the multicomponent counts posterior was constructed, we choose the same low-ranking statistic threshold when counting the number of recovered injections. However, it is not trivial to map a given trigger to an injection, and count that trigger as a recovered injection, given the low ranking statistic threshold used and the resulting preponderance of background triggers. We therefore propose the following method. (Note that a similar method, using the two-component counts posterior, was used for the determination of $\langle VT \rangle$, in \cite{GW170104-DETECTION}.) 

Let $[x_1,...,x_N]$ be the list of triggers produced during a search, and let $[\gamma_1,\gamma_2,...,\gamma_I]$ be the list of triggers produced during the injection campaign. Suppose now we include one trigger from the injection campaign, to the list of triggers from the search. We now have the following list $[x_1, x_2, ..., x_N, \gamma_i]$. The contribution of $\gamma_i$ to the mean value of the count for category $\alpha$ can be computed using Eq.~\ref{mean_diff}. We also compute the contributions to the mean values pertaining to all the other astrophysical source categories, due to $\gamma_i$, and determine an aggregated contribution:
\begin{widetext}
\begin{equation}
\Delta\Lambda_{\gamma_i} = \sum_{\alpha}\frac{\mathrm{cov}_N(\Lambda_0, \Lambda_{\alpha}) + \sum_{\beta}\mathrm{cov}_N(\Lambda_{\alpha},\Lambda_{\beta})K_{\beta}(\gamma_i)}{\langle \Lambda_0 \rangle_N + \sum_{\beta}K_{\beta}(\gamma_i)\langle \Lambda_{\beta} \rangle_N}
\end{equation}
\end{widetext}
where both $\alpha$ and $\beta$ are summed over astrophysical source categories, and $\mathrm{cov_N}$ is defined in Appendix \ref{sec:appB}.

We repeat this procedure for all triggers from the injection campaign, adding each one separately (with replacement) to the triggers from the search, and determining the change in the mean due to each addition. $N_{\mathrm{rec}}$ is estimated as the sum of the contributions to the mean due to all $\gamma_i$s in $[\gamma_1,\gamma_2,...,\gamma_I]$:
\begin{equation}\label{nrec}
N_{\mathrm{rec}} = \sum_{i=1}^{I}\Delta\Lambda_{\gamma_i}
\end{equation}

\section{Rates Posterior with Uncertainties}\label{sec:uncertainty}

As such, going from the counts posterior to the rates posterior is a trivial change of variables. Thought of in another way, rates and counts are essentially the same quantity expressed in different units:
\begin{equation}\label{rate_ratio}
\vec{R} = \left\lbrace \frac{\Lambda_{\alpha}}{(VT)_{\alpha}} \right\rbrace, \alpha = 1,2,...,Q
\end{equation}
where, as before, $Q$ is the total number of astrophysical components. 

There are, however, uncertainties associated with the determination of the spacetime volume sensitivity, arising from calibration errors when measuring the detector strain $h$ \footnote{The uncertainty in the GW amplitude $h$ measured from detector data gets translated to an uncertainty in the detector range $d$ at leading order ($d \sim h$), or equivalently, an uncertainty in volume ($V \sim h^3 \Rightarrow \frac{dV}{V} \sim 3\frac{dh}{h}$);(see \cite{GW150914-RATES-SUPPLEMENT}, specifically section 5 therein.)}:
\begin{equation}
S_{cal} \approx 3\frac{\Delta h}{h}
\end{equation}
as well as statistical Monte Carlo errors when evaluating Eq. \ref{VT_integral}:
\begin{equation}
S_{stat} = \frac{1}{\sqrt{N_{rec}}}.
\end{equation}
Let $S$ be the total fractional error associated with these uncertainties:
\begin{equation}
S = \sqrt{S^2_{stat} + S^2_{cal}}.
\end{equation}
One can then model the measurement distribution on $VT$ as a log-normal: 
\begin{widetext}
\begin{equation}\label{lognormal}
p(VT~|~\langle VT \rangle, S) = \frac{1}{VT\cdot S\sqrt{2\pi}}\exp\left[-\frac{1}{2}\left(\frac{\ln VT -\ln\langle VT \rangle}{S}\right)^2\right]
\end{equation}
\end{widetext}
where $\langle VT \rangle$ is assumed to be the same as the measured population-averaged spacetime volume in Eq.~\eqref{VT_integral}.

There are now two ways in which the distributions on the Poisson counts, and the distribution on the spacetime volume sensitivity, can be combined to evaluate the posterior on the rates. 

The first method is a direct application of the ratio distribution: given two positive random variables, $y_1$ and $y_2$, with joint distribution $f(y_1,y_2)$, the distribution on the ratio of these two variables goes as:
\begin{equation}
p(u \equiv y_2/y_1) = \int_0^{\infty}y_1f(y_1, uy_1)dy_1.
\end{equation}
Using the ratio in Eq. \eqref{rate_ratio}, we identify $y_1$ as $VT$ and $y_2$ as $\Lambda$, with $y_1$ distributed as given in Eq. \eqref{lognormal} and $y_2$ distributed as given in Eq. \eqref{counts_posterior}. And assuming the joint distribution $f(y_1,y_2)$ is the product of Eqns.\eqref{lognormal} and \eqref{counts_posterior}, we can put down a distribution on the rate:
\begin{equation}\label{calibration}
p(R_{\alpha}~|~\vec{x}, \langle VT \rangle_{\alpha}, S_{\alpha}) = \frac{1}{S_{\alpha}\sqrt{2\pi}}\int_{-\infty}^{\infty}p(R_{\alpha}\cdot e^{v_{\alpha}}~|~\vec{x}, \langle VT \rangle_{\alpha})\exp\left[-\frac{1}{2}\left(\frac{v_{\alpha}}{S_{\alpha}}\right)^2 + v_{\alpha} \right]dv_{\alpha}
\end{equation}
where $v_{\alpha} = \ln\left(\frac{(VT)_{\alpha}}{\langle VT \rangle_{\alpha}}\right)$, and $p(R_\alpha \cdot e^{\nu_{\alpha}}~|~\vec{x}, \langle VT \rangle_{\alpha})$ is the marginalized counts posterior (cf. Eq.\ref{counts_posterior})  for source-category $\alpha$ with the change of variable $\Lambda_\alpha \rightarrow \Lambda_\alpha/<VT>_\alpha = R_\alpha e^{\nu_\alpha}$, given a measured spacetime volume sensitivity $\langle VT \rangle_{\alpha}$). 

The second method starts by asserting that $R$ and $VT$ are independent random variables, and the joint probability distribution on these two variables (we have dropped the subscript $\alpha$ for notational simplicity):
\begin{equation}
p(R, VT~|~\langle VT \rangle, S) = p(R)\cdot p(VT~|~\langle VT \rangle, S) 
\end{equation}
acts as the prior for the joint posterior on $R$ and $VT$:
\begin{equation}
p(R, VT~|~\vec{x}, \langle VT \rangle, S) \propto p(R, VT~|~\langle VT \rangle, S) \cdot p(\vec{x}~|~R, VT).
\end{equation}
Here, $p(\vec{x}~|~R, VT)$ is the FGMC likelihood, and $p(VT~|~\langle VT \rangle, S)$ is modeled with the log-normal distribution as in Eq.~\eqref{lognormal}. Marginalizing out $VT$ from the above equation gives the sought after rate posterior. Assuming further that the prior on the rate $p(R)$ follows a power law, the rate posterior becomes up to a normalization constant (re-introducing the subscript $\alpha$ to simplify comparing the equation below with \eqref{calibration}):
\begin{equation}
p(R_{\alpha}~|~\vec{x}, \langle VT \rangle_{\alpha}, S_{\alpha}) \propto \int_{-\infty}^{+\infty}p(R_{\alpha}\cdot e^{v_{\alpha}}~|~\vec{x}, \langle VT \rangle_{\alpha}~)\exp\left[-\frac{1}{2}\left(\frac{v_{\alpha}}{S_{\alpha}}\right)^2 -av_{\alpha} \right]dv_{\alpha}
\end{equation}
where $a$ is the exponent of the power-law priors on $R$. The rate posteriors in the catalog either use a uniform prior ($a = 0$) for categories with no confirmed detections (viz. NSBH), and a Jeffreys prior ($a = -0.5$) for the other categories (viz. BNS, BBH, Terrestrial). 

\section{Illustrative Results}\label{sec:results}

As a proof of principle, we apply this multi-component extension of the FGMC method to a mixture of synthetic BNS, NSBH and BBH signals added to real detector data from O1 and O2 devoid of astrophysical GW signals. Making the data free from GW signals is achieved by a method referred to as ``time-sliding'': the detector strain time series from one of the detectors is translated in time, with respect to the strain time series in another detector, by an amount greater than the light-travel time between the detectors; coincident events post this time-shifting are then used.

The BNS, NSBH and BBH signals were injected into O1 and O2 detector data, and recovered with the GstLAL pipeline, separately. From those injections that were recovered, as determined by whether a trigger exists within a 1 second time window of the injections, 30 BNS injections, 30 NSBH injections, and 100 BBH injections were selected at random. Their corresponding triggers were then added to the list of triggers produced when analyzing the time-slid data.   

The set of CBC signals used for the injection campaigns were selected as follows: 

BNS signals were drawn at random from a ``broad'' distribution of synthetic BNS signals.  The component masses $m_i$ were distributed uniformly between $0.8 \leq m_i/M_{\odot} \leq 2.3$. The component spins were isotropically distributed on the sphere with a maximum spin magnitude of $0.4$. 

NSBH signals were drawn at random from each of three delta-function distributions of synthetic NSBH signals. The delta-functions were centered at component masses $(1.4M_{\odot}, 5M_{\odot})$ (low-mass NSBHs), $(1.4M_{\odot}, 10M_{\odot})$, $(1.4M_{\odot}, 30M_{\odot})$ (high-mass NSBHs), and the spins for each of these sets were aligned with the orbital angular momentum of the binary, with a maximum allowed spin magnitude of $0.05$ for the neutron star component, and $0.99$ for the black hole component.

BBH signals were drawn at random from a ``broad'' distribution of synthetic BBH signals. The component masses $m_i$ were distributed uniformly in $\ln m_i$, between $5 \leq m_i/M_{\odot} \leq 100$, and a total mass cutoff at $100 M_{\odot}$. The component spins were aligned with the orbital angular momentum of the binary, with a maximum allowed spin magnitude of $0.99$.

All sets from which the various synthetic signals were drawn, ensured that the binary systems were distributed uniformly in co-moving volume.

\begin{figure*}
        \centering
        \begin{subfigure}[b]{0.475\textwidth}
            \centering
            \includegraphics[width=\textwidth]{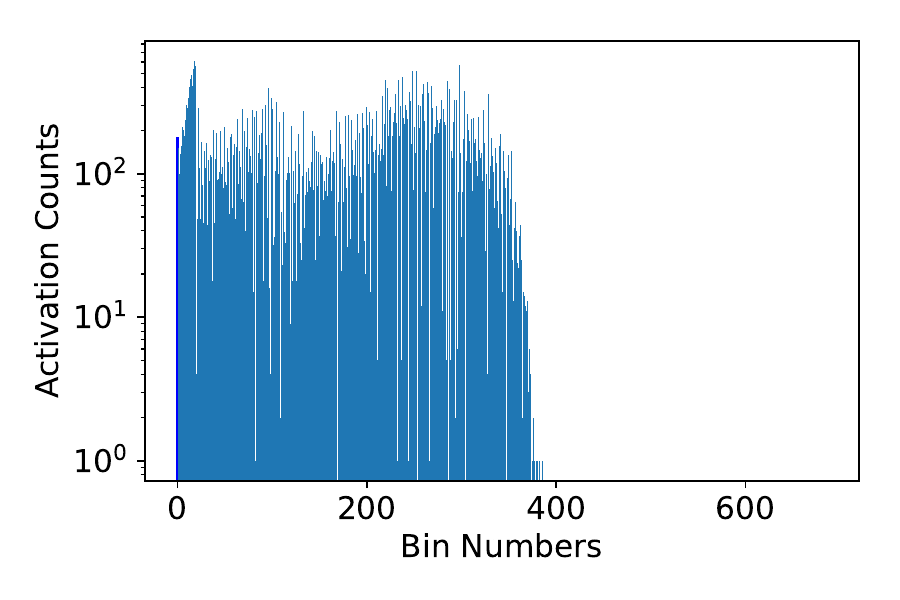}
            \caption[BNS]%
            {{\small BNS}}    
            \label{fig:bns_activation}
        \end{subfigure}
        \hfill
        \begin{subfigure}[b]{0.475\textwidth}  
            \centering 
            \includegraphics[width=\textwidth]{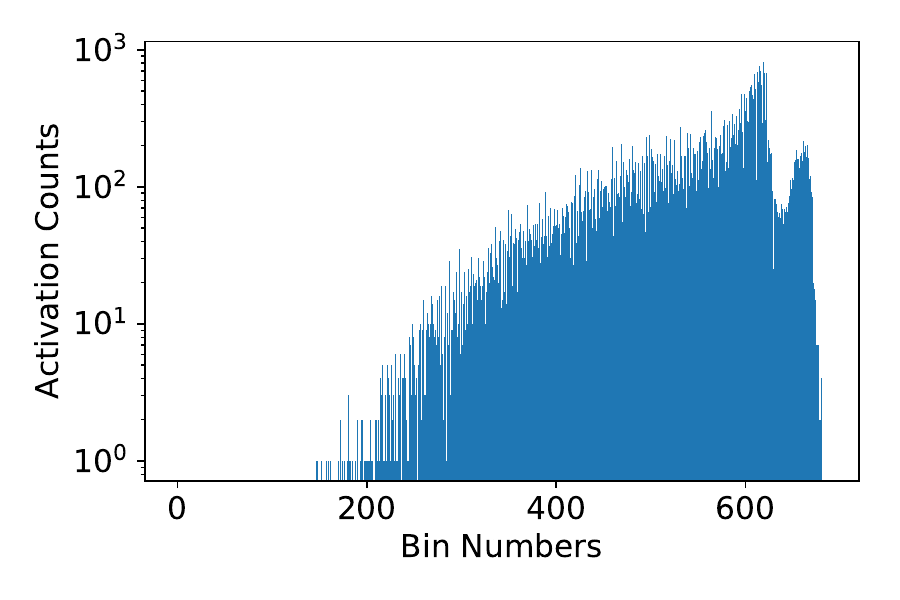}
            \caption[]%
            {{\small NSBH}}    
            \label{fig:nsbh_activation}
        \end{subfigure}
        \vskip\baselineskip
        \begin{subfigure}[b]{0.475\textwidth}   
            \centering 
            \includegraphics[width=\textwidth]{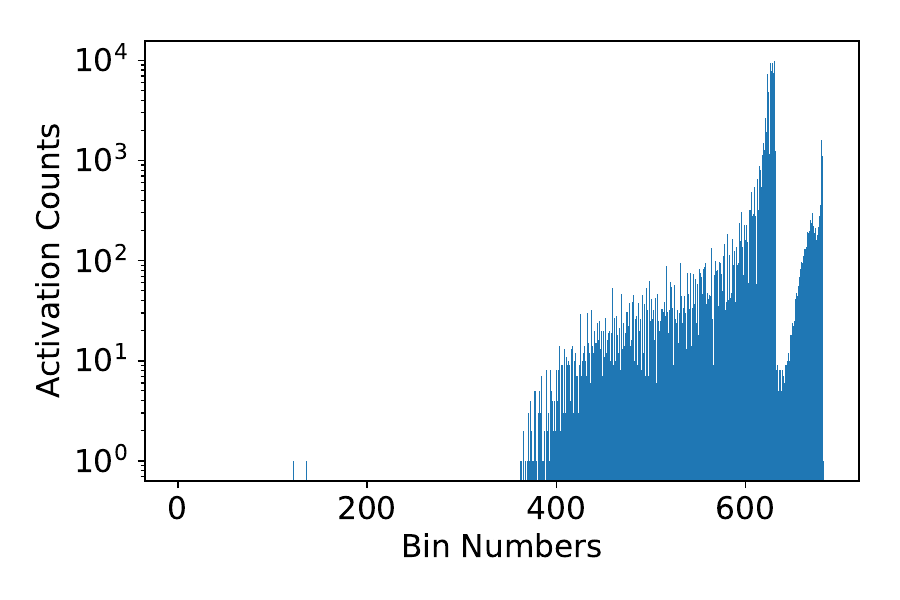}
            \caption[]%
            {{\small BBH}}    
            \label{fig:bbh_activation}
        \end{subfigure}
        \hfill
        \begin{subfigure}[b]{0.475\textwidth}   
            \centering 
            \includegraphics[width=\textwidth]{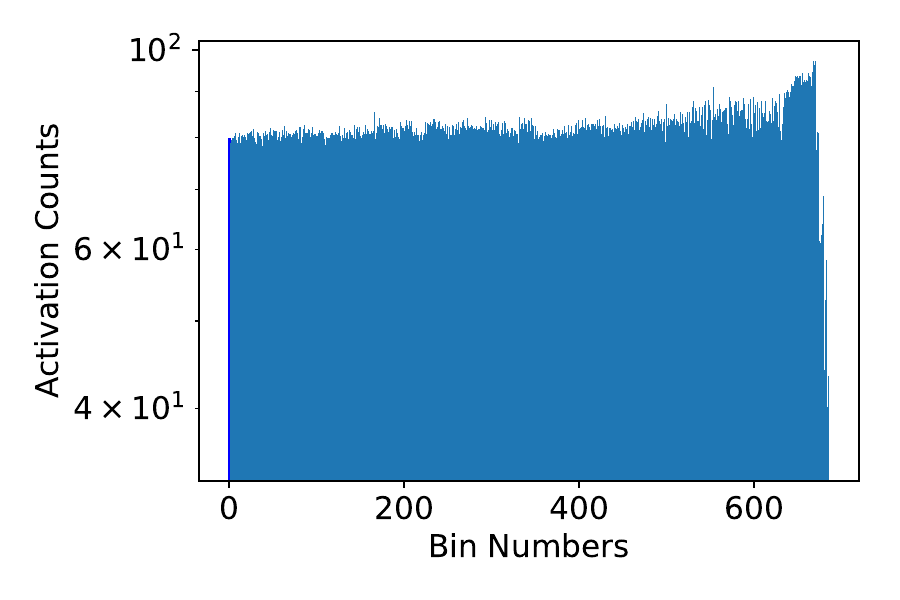}
            \caption[]%
            {{\small Background}}    
            \label{fig:background_activation}
        \end{subfigure}
        \caption[ The average and standard deviation of critical parameters ]
        {\small The panels in this figure show how often each of the 686 non-overlapping regions of the template bank, denoted by a bin number, get ``rung up'' when recovering BNS, NSBH, and BBH signals. An estimate of how background events distribute themselves in the template bank is also given; this estimate is evaluated by GstLAL when analyzing real detector data devoid of synthetic signals. The weights $W_{\alpha}(m)$, $\alpha = \lbrace$ BNS, NSBH, BBH, 0 $\rbrace$, $m = \lbrace 0, 1, \ldots, 685 \rbrace$ (cf. Eq.~\eqref{weights}) are determined from these histograms.}
        \label{fig:activation_counts}
    \end{figure*}

Employing the GstLAL detection pipeline, this time-slid data added with synthetic signals was analyzed and candidate events were assigned foreground ($p(L|m,\mathrm{signal})$) and background ($p(L|m,\mathrm{noise})$) distribution values. Furthermore, using software injection campaigns described in section \ref{template_weights}, bin-dependent template weights $W_{\alpha}(m)$ ($\lbrace \alpha = \mathrm{BNS, NSBH, BBH} \rbrace $) were estimated (see Equation \eqref{weights} and Figure~~\ref{fig:activation_counts}).

\begin{figure}
\centering
\includegraphics[scale=0.50]{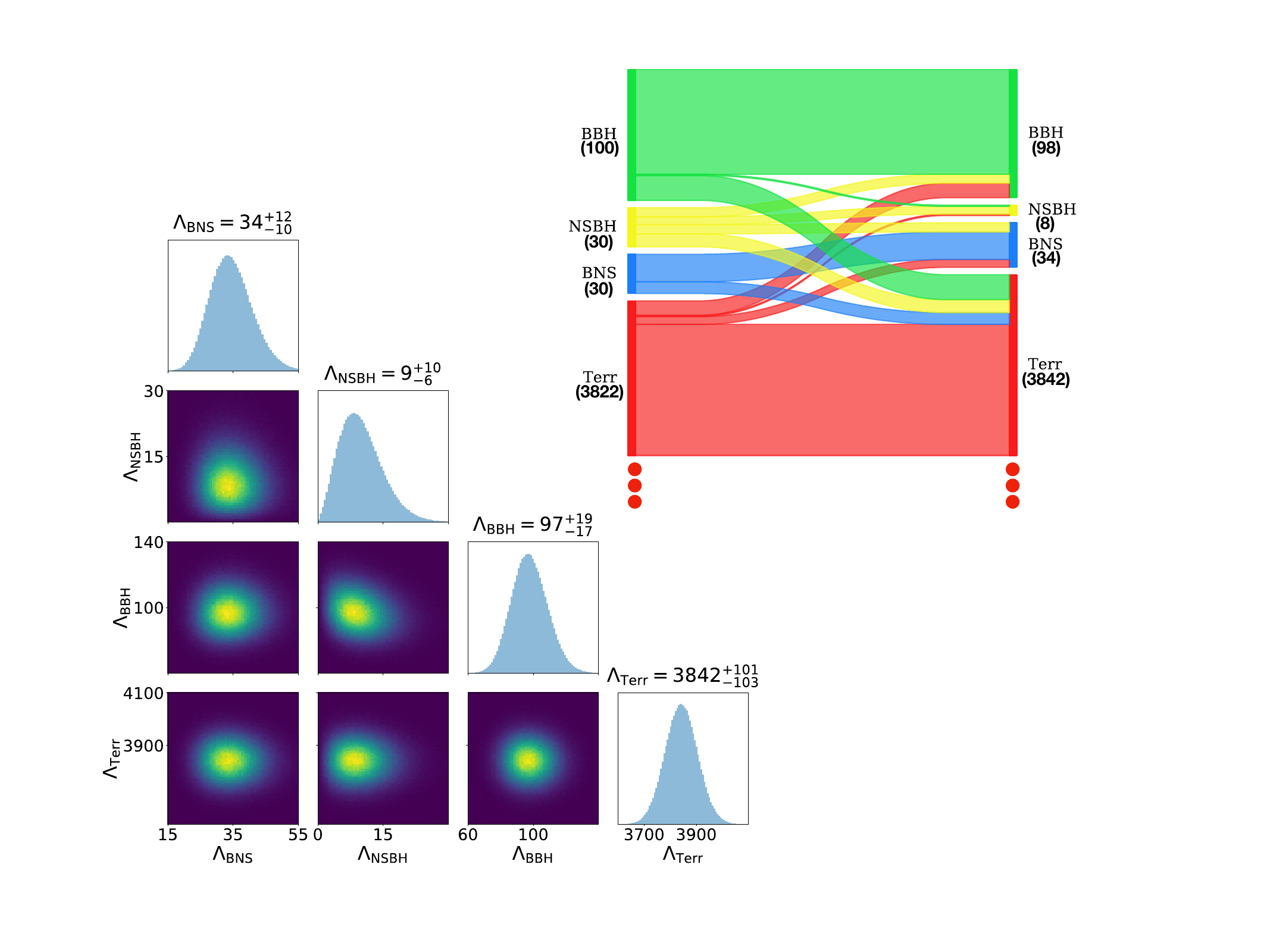}
\caption{The {\bf corner plot} provides a visual representation of the multi-component posterior on the Poisson counts for BNS, NSBH, BBH, Terr (Background/Terrestrial) events; this plot also contains the median values (with $90\%$ symmetric confidence intervals as error bars) of the Poisson counts. Time-slid O1 and O2 data, injected with 30 BNS, 30 NSBH and 100 BBH signals, were used. The median BNS and and BBH counts are approximately equal to the numbers of BNS and BBH signals injected in the data, and these numbers are well within the error bars. However, the posterior on the NSBH has a median value much lower than the number of injected NSBH signals. This results from a combination of factors: low-mass NSBH signals being partially confused as BNS signals, high mass NSBH signals being partially confused as BBH signals, and NSBH signals that were not recovered by the pipeline with high significance being confused as background. The {\bf Sankey diagram} - with the number of injected signals depicted on the left hand side, and the aggregated categorical astrophysical probabilities on the right hand side - also illustrates this. It further reveals that the few low-significance BNS and BBH signals classified as noise are being compensated for by high-significance NSBH signals and terrestrial events counted as BNS and BBH signals, resulting in BNS and BBH signals being counted as their appropriate categories on average. }
\label{fig:counts_corner_plot}
\end{figure}

We subsequently computed the Bayes factors $\vec{K}_{\alpha}(x)$ and constructed the multi-component joint posterior on the Poisson counts for each of the categories. Via a Python implementation of Markov-Chain-Monte-Carlo described in \cite{emcee}, we sampled the multi-component posterior. The corner plot in Figure~\ref{fig:counts_corner_plot}, and the median values (with a $90\%$ symmetric confidence interval as error bar) of the Poisson counts reveal that while most BNS and BBH signals are being categorized, on average, as BNS and BBH signals, respectively, some of the NSBH signals are being at least partially confused as BNS or BBH signals. This is perhaps not so surprising if Figure \ref{fig:counts_corner_plot} is viewed in light of Figure \ref{fig:activation_counts}. The set of bins in the template bank activated by BNS signals is not disjoint from the set activated by NSBH signals. The same is also true for NSBH and BBH signals. We would therefore expect to be correlation between BNS and NSBH categories, and NSBH and BBH categories, resulting in signals being partially counted in the wrong category.

We also computed categorical astrophysical probabilities (cf. section~\ref{sec:pastro}) for each of the $\sim 4000$ triggers in the data set. Rather than tabulate the results for this big list, we depict, via the Sankey diagram in Figure~\ref{fig:counts_corner_plot}, how the multi-component FGMC formalism is categorizing and counting these triggers. A priori, we know which ones of these triggers correspond to a time-slide event, and which ones correspond to an injection. The thickness of the bands on the left hand side of the Sankey diagram is proportional to these numbers. The aggregated astrophysical probabilities, by source category, are also represented by a proportionately thick band on the right hand side of the Sankey diagram. Figure~\ref{fig:counts_corner_plot} shows that most BBHs and BNSs are counted as BBHs and BNSs, respectively, and there is no confusion between these two categories. This can again be understood with the help of Figure~\ref{fig:activation_counts}: the regions of the template bank activated by BNS and BBH signals are almost disjoint. On the other hand, the overlap in the regions activated by BNS and NSBH signals, and BBH and NSBH signals, causes low-mass NSBH signals to be partially counted as BNS signals, and high-mass NSBH signals to be partially counted as BBH signals. There is also some correlation between the Terrestrial category and the astrophysical categories, resulting in weaker signals being partially counted as noise, and vice-versa. These misclassifications are discussed further in the next section.

Note that the aggregated categorical astrophysical probabilities are approximately equal to the mean values of the Poisson counts \footnote{Equation \ref{mean_pastro} suggests that the choice of prior becomes increasingly irrelevant as $\langle \Lambda_{\alpha} \rangle$ becomes sufficiently large. This is simply an indication that, when we've accumulated many events of category $\alpha$, the counts posterior is informed primarily by the data, and not any prior assumption on the distribution of the counts.}:
\begin{equation}\label{mean_pastro}
\sum_{j=1}^{N} P_{\alpha}(x_j~|~\vec{x}) = \langle \Lambda_{\alpha} \rangle - a - 1
\end{equation}
where, like in section~\ref{sec:uncertainty}, $a$ is the exponent on the Poisson count for category $\alpha$ in the joint prior for the multi-component FGMC posterior, assuming that the prior has the form $p(\Lambda_0, \vec{\Lambda}_1) = p(\Lambda_0,\Lambda_{\alpha}\notin\vec{\Lambda}_1) \Lambda_{\alpha}^a$ (see appendix~\ref{sec:appD}). An identical formula also exists connecting the terrestrial probabilities with the mean of the terrestrial Poisson count. Thus, the right-hand side values (aggregated astrophysical and terrestrial probabilities) in the Sankey diagram of Figure~\ref{fig:counts_corner_plot} may be thought of as mean values on the Poisson counts.

\section{Summary and Outlook}\label{sec:conclusion}

Inferring rates of compact binary mergers from gravitational-wave data is an important science goal of the LIGO-Virgo Collaboration (LVC). GWTC-1 \cite{GWTC-1} (or the ``catalog'', as we've been calling it in this paper), quoted rate estimates for the mergers of BNSs, NSBHs, and BBHs, using GW data from the LIGO-Virgo network of ground-based interferometric detectors, and analyzed by two separate matched-filter based pipelines (GstLAL and PyCBC). This paper serves as a supplement to the catalog, by describing the formalism used to produce rate estimates from data analyzed by GstLAL. The formalism itself is inspired by a Poisson-statistics based counting method developed by W. Farr, J. Gair, I. Mandel, and C. Cutler (abbreviated to ``FGMC'' in this paper) in the context of GW data analysis \cite{FGMC} (although similar work in non-GW contexts was done earlier, e.g: \cite{Loredo-Wasserman}). 

Assuming that candidate events triggered by terrestrial phenomena or astrophysical GWs occur as independent Poisson processes, the original FGMC formalism constructs a joint posterior conditional probability distribution from the significance of triggers assigned by a detection pipeline (e.g: GstLAL), as measured by the Bayes-Factor (see Section~\ref{sec:posterior}), above a low-ranking statistic threshold, on the Poisson expected counts for astrophysical and terrestrial CBC events. 

The multi-component extension of the FGMC formalism delineated in this paper constructs, from the foreground distribution of ranking statistics estimated by GstLAL, an arbitrary number of foreground probability distributions, each targeted at a specific astrophysical source class. This is achieved by a mass-based template-weighting method. The weights are determined via injection campaigns, by studying the distribution of templates, across the template bank, that got ``rung-up'' during the recovery of injections.

A proof-of-principle application of this multi-component extension was conducted on time-slid O1 and O2 data, added with synthetic BNS, NSBH and BBH signals, and analyzed with the GstLAL detection pipeline. We find that BNS and BBH signals are for the most part counted appropriately as BNS and BBH. On the other hand, there exist correlations between low-mass NSBH signals and BNS signals, as well as between high mass NSBH signals, and BBH signals, causing partial mis-categorization of NSBH signals. There are also correlations between the astrophysical categories, and the terrestrial category.

Is the confusion in counting and labeling of triggers a cause for concern? 
It is worth pointing out here that the partial mis-classification of the astrophysical signals as terrestrial signals, and vice-versa, is not a limitation of this method per se, but rather a consequence of the significance assigned to these events by the detection pipeline. Nevertheless,from the perspective of computing rates, the partial mis-classification of astrophysical and terrestrial signals is somewhat compensated for by the fact that the same threshold on the ranking statistic is used when determining the counts posterior, and the spacetime volume sensitivity. The {\it ratio} of these two quantities, which gives the astrophysical rate, is largely unaffected by this choice of threshold, provided the threshold ensures a preponderance of terrestrial triggers \cite{FGMC}.

However, the confusion between low (high) mass NSBH signals and BNS (BBH) signals, would be a concern, if this significantly changed the counts of these types of signals. Nevertheless, the extensive follow-up analysis using the parameter estimation given in Sec. V and Table III of the catalog show that there were no NSBH signals of any sort detected. Thus, the misclassification of NSBH events at the border between source categories, is not a concern for the data set being analyzed, viz. O1 and O2 data. 

Parameter estimation studies show that 10 significant BBH detections and 1 significant BNS detection, were uncovered from the data. These findings are consistent with the multi-component rates results given in the catalog and computed using the methods described in this paper. The multi-component rates are also consistent with separate (albeit simpler) rates analyses described in the catalog, which serves as a nice sanity check for the method described in this paper. 

However, we will need to extend the methods described here in the future to better identify astrophysical signals at the boundaries between source-categories. For example, a computationally intensive method involving a synergy of the formalism delineated in this paper and the samples provided by accurate parameter estimation studies, such as the one described in \cite{FGDV}, might be required.

It is worth noting here that the application of the multi-component extension of FGMC used in the catalog is by no means unique. One need not restrict oneself to just BNSs, NSBHs, and BBHs. Additional astrophysical source classes could be added, like, for example, CBCs from the putative ``mass-gaps'' (with binary component mass range $3M_{\odot}-5M_{\odot}$ and above $45 M_{\odot}$ \cite{O2:RP}), and different categories of black holes. Moreover, the redistribution of foreground events need not involve mass-based weighting alone; for example, work on constructing redshift dependent foreground distributions, to study redshift dependent rates of BBHs, is expected to be published soon \cite{chase:in-prep}.

\section{Acknowledgements} \label{sec:ack}

We thank the LIGO-Virgo Scientific Collaboration for access to data. LIGO was
constructed by the California Institute of Technology and Massachusetts
Institute of Technology with funding from the National Science Foundation (NSF)
and operates under cooperative agreement PHY-0757058. We gratefully acknowledge the support by NSF grant PHY-1626190 for the UWM computer cluster. We also thank Deep Chatterjee, Shaon Ghosh and Chad Hanna for illuminating discussions. SJK gratefully acknowledges suppport through NSF grant PHY-1607585. SRM thanks the LSSTC Data Science Fellowship Program, which is funded by LSSTC, NSF Cybertraining Grant $\#$1829740, the Brinson Foundation, and the Moore Foundation.

%\bibliography{references}

%%%%%%%%%%%%%%
% APPENDICES
%%%%%%%%%%%%%%
\appendix
\section{Derivation of the multicomponent FGMC counts posterior}\label{sec:appA}

The derivation follows \cite{Loredo-Wasserman, FGMC}.

Consider a time interval T during which gravitational-wave (GW) observations are made. We divide this interval into $N_t$ time-fragments:
\begin{equation}
\delta t = \frac{T}{N_t}.
\end{equation}
We make $N_t$ large enough to ensure that at most 1 event occurs over time interval $\delta t$.

Let us assume that there are N intervals containing exactly $1$ event and $N_t-N$ intervals containing exactly $0$ events. We denote by $x$ the properties of an event. Previously (i.e, in \cite{FGMC}), this was simply the ranking statistic. Here, $x \rightarrow \lbrace{L, m \rbrace}$, where $L$ is the ranking statistic and $m$ is a set of template parameters. Thus, $x_i$ is the only trigger in the ith time interval, with detection statistic value $L_i$ and template parameters $m_i$. We denote by $\varnothing$ a time interval that contains no events. Thus, $\varnothing_j$ is the jth time interval that contains no events. 

%If $\Lambda_0$ and $\Lambda_\alpha$ are the Poisson expected counts for a time interval $T$, then the probability of there being only $1$ trigger $x_i$ in the ith time interval is:
%\begin{equation}
%p(x_i | \Lambda_0, \vec{\Lambda}_1) = p(L_i, m_i | \Lambda_0, \vec{\Lambda}_1) 
%\end{equation} 
%where $\Lambda_0$ is the Poisson count for triggers of terrestrial origin, and $\vec{\Lambda}_1 = \lbrace{\Lambda_\alpha, \alpha = \mathrm{BNS}, \mathrm{NSBH}, \ldots \rbrace}$ is the Poisson count for triggers of astrophysical origin.
%
%On the other hand, the probability of there being no trigger in the jth time interval is: $p(\varnothing_j | \Lambda_0, \vec{\Lambda})$

The joint probability of these N propositions gives us the likelihood up to a combinatorial constant:
\begin{equation}
\mathcal{L} \propto \prod_{i = 1}^N p(x_i | \Lambda_0, \vec{\Lambda}_1)\times\prod_{j=N+1}^{N_t} p(\varnothing_j | \Lambda_0, \vec{\Lambda}_1).
\end{equation}

Computing $p(\varnothing_j | \Lambda_0, \vec{\Lambda}_1)$ is a straightforward application of the Poisson distribution for $0$ counts in a time interval $\delta t$:
\begin{equation}
p(\varnothing_j | \Lambda_0, \vec{\Lambda}_1) \propto \exp(-\lambda).
\end{equation}
where $\lambda$ is the expected number of counts (terrestrial and astrophysical combined) in interval $\delta t$.
Meanwhile:
\begin{equation}
p(x_i | \Lambda_0, \vec{\Lambda}_1) = p(x_i | 1, \Lambda_0, \vec{\Lambda}_1)p(1 | \Lambda_0, \vec{\Lambda}_1).
\end{equation}
where $p(1 | \Lambda_0, \vec{\Lambda}_1)$ is the probability of acquiring exactly 1 event in an interval of duration $\delta t$, which in turn can be computed via straightforward application of the Poisson distribution:
\begin{equation}
p(1 | \Lambda_0, \vec{\Lambda}_1) = \lambda\exp(-\lambda)
\end{equation}
and $\lambda$, the mean number of triggers in interval $\delta t$ is:
\begin{equation}
\lambda = \frac{1}{N_t}\cdot \left(\Lambda_0 + \sum_\alpha \Lambda_\alpha \right).
\end{equation}
On the other hand, $p(x_i|1,\Lambda,\vec{\Lambda}_1)$ is the fraction of triggers with detection statistic $L_i$ in bin $m_i$,
\begin{equation}
p(x_i~|~1,\Lambda_0,\vec{\Lambda}_1) = \frac{\Lambda_0 p(L_i,m_i | \mathrm{noise}) + \sum_{\alpha} \Lambda_{\alpha} p(L_i, m_i~|~{\alpha})}{\Lambda_0 + \sum_{\alpha} \Lambda_{\alpha}}.
\end{equation}
Therefore, the likelihood becomes:
\begin{widetext}
\begin{equation}
\mathcal{L} \propto \exp\left(-\left(\Lambda_0 + \sum_\alpha \Lambda_\alpha \right) \right)\cdot \prod_{i = 1}^N\left(\Lambda_0 \cdot p(L_i,m_i|\mathrm{noise}) + \sum_\alpha \Lambda_\alpha \cdot p(L_i, m_i | \alpha)\right).
\end{equation}
\end{widetext}
The multicomponent counts posterior, up to a normalization constant, thus has the following general form:
\begin{widetext}
\begin{equation}\label{counts_posterior_der}
p\left(\Lambda_0, \vec{\Lambda}_1~|~\vec{x}~\right) = p\left(\Lambda_0, \vec{\Lambda}_1 \right)\cdot \exp\left(-\left(\Lambda_0 + \sum_\alpha \Lambda_\alpha \right) \right)\cdot\prod_{i = 1}^N\left(\Lambda_0 \cdot p(L_i,m_i|\mathrm{noise}) + \sum_\alpha \Lambda_\alpha \cdot p(L_i, m_i | \alpha)\right).
\end{equation}
\end{widetext}
The distributions (normalized density functions), $p(L_i,m_i |\mathrm{noise})$ and $p(L_i,m_i |\mathrm{\alpha})$ in Equation \ref{counts_posterior_der} can be divided into two pieces each:
\begin{widetext}
\begin{eqnarray}
p(L_i, m_i~|~\mathrm{noise}) &=& p(L_i~|~m_i, \mathrm{noise})\cdot W_0(m_i) \\
p(L_i, m_i~|~\alpha) &=& p(L_i~|~m_i, \alpha)\cdot p(m_i~|~\alpha) \approx p(L_i~|~m_i, \mathrm{signal})\cdot W_{\alpha}(m_i).
\end{eqnarray}
\end{widetext}
where the weights $W$ are defined as:
\begin{eqnarray}
W_0(m_i) \equiv p(m_i~|~\mathrm{noise}), \\
W_{\alpha}(m_i) \equiv p(m_i~|~\alpha).
\end{eqnarray}
$p(L_i~|~m_i,\alpha) \approx p(L_i~|~m_i,\mathrm{signal})$ is a statement of the universality of the ranking statistic distribution $L$ under the foreground model \cite{GW150914-RATES-SUPPLEMENT, ChenAndHolz, Schutz2011}; in other words, the foreground model is not expected to change appreciably for different classes of astrophysical signals. 

\section{Evolution of the Counts posterior with the addition of candidate events}\label{sec:appB}

\subsection{Updating the Multicomponent counts posterior}

Having constructed the multi-component counts posterior, it is useful to investigate how the posterior evolves with the addition of candidate events. 

Let $p_{N}(\vec{\Lambda}_1~ | ~ \vec{x})$ be the counts posterior constructed from a set of N candidate events, appropriately normalized. Suppose we now acquire an additional trigger, $x_{N+1}$. We wish to determine how the inclusion of this trigger modifies the counts posterior. It is straightforward to see, from equation \ref{counts_posterior}, that:
\begin{equation}
p_{N +1}(\Lambda_0, \vec{\Lambda}_1~|~\vec{x}) = A\cdot p_{N}(\vec{\Lambda}_1~|~\vec{x})\cdot \left[\Lambda_0  + \vec{\Lambda}_1\cdot \vec{K}(x_{N+1})\right] 
\end{equation}
where $A$ is a constant that we can determine via normalization:
\begin{equation}\label{normalization}
\int_{0}^{\infty} p_{N +1}(\Lambda_0, \vec{\Lambda}_1~|~\vec{x})d\Lambda_0d\vec{\Lambda}_1 = 1.
\end{equation}
Writing $p_{N +1}(\Lambda_0, \vec{\Lambda}_1~|~\vec{x})$ in terms of $p_{N}(\vec{\Lambda}_1 | \vec{x})$ and carrying out the above integral yields:
\begin{widetext}
\begin{eqnarray}
\int_{0}^{\infty} p_{N +1}(\Lambda_0, \vec{\Lambda}_1~|~\vec{x})d\Lambda_0 d\vec{\Lambda}_1 &=& \int_0^{\infty} A\cdot p_{N}(\Lambda_0, \vec{\Lambda}_1~|~\vec{x})\cdot \left[\Lambda_0 + \vec{\Lambda}_1\cdot \vec{K}(x_{N+1})\right]d\Lambda_0d\vec{\Lambda}_1 \\
&=& A\left[\langle \Lambda_0 \rangle_N + \sum_{\alpha}K_{\alpha}(x_{N+1})\langle \Lambda_{\alpha} \rangle_N \right]
\end{eqnarray}
\end{widetext}
where the quantities in the angular brackets are the mean values of the counts (whether of terrestrial or astrophysical origin), defined as:
\begin{equation}\label{mean}
\langle \Lambda_{\beta} \rangle_N \equiv \int_0^{\infty} \Lambda_{\beta} p_N(\Lambda_0, \vec{\Lambda}_1)d\Lambda_0d\vec{\Lambda}_1.
\end{equation}
Thus, from equation \ref{normalization}:
\begin{equation}
A(x_{N+1}) = \frac{1}{\langle \Lambda_0 \rangle_N + \sum_{\alpha}K_{\alpha}(x_{N+1})\langle \Lambda_{\alpha} \rangle_N}. 
\end{equation}
Therefore, the multicomponent counts posterior is updated to be:
\begin{widetext}
\begin{equation}\label{updated_posterior}
p_{N +1}(\Lambda_0, \vec{\Lambda}_1~|~\vec{x}) = p_{N}(\Lambda_0, \vec{\Lambda}_1~|~\vec{x})\cdot\frac{\Lambda_0  + \vec{\Lambda}_1\cdot \vec{K}(x_{N+1})}{\langle \Lambda_0 \rangle_N + \sum_{\alpha}K_{\alpha}(x_{N+1})\langle \Lambda_{\alpha} \rangle_N}.
\end{equation}
\end{widetext}
\subsection{Updating the mean value of the counts}
The change in the posterior due to the addition of an event leads to changes in the mean values of the Poisson expected counts for each source category.

From equations \ref{mean} and \ref{updated_posterior}:
\begin{widetext}
\begin{eqnarray}
\langle \Lambda_{\beta} \rangle_{N+1} &\equiv & \int_0^{\infty} \Lambda_{\beta} p_{N+1}(\Lambda_0, \vec{\Lambda}_1~|~\vec{x}_{N+1})d\Lambda_0d\vec{\Lambda}_1 \\
& = & \frac{\langle \Lambda_0 \Lambda_{\beta} \rangle_N + \sum_{\alpha} \langle \Lambda_{\alpha}\Lambda_{\beta} \rangle_N K_{\alpha}(x_{N+1})}{\langle \Lambda_0 \rangle_N + \sum_{\alpha}K_{\alpha}(x_{N+1})\langle \Lambda_{\alpha} \rangle_N}.
\end{eqnarray}
\end{widetext}
The change in the mean value due to the addition of the $N+1$th trigger is:
\begin{widetext}
\begin{eqnarray}\label{mean_diff}
\langle \Lambda_{\beta} \rangle_{N+1} - \langle \Lambda_{\beta} \rangle_{N} &=& \frac{\langle \Lambda_0 \Lambda_{\beta} \rangle_N + \sum_{\alpha} \langle \Lambda_{\alpha}\Lambda_{\beta} \rangle_N K_{\alpha}(x_{N+1})}{\langle \Lambda_0 \rangle_N + \sum_{\alpha}K_{\alpha}(x_{N+1})\langle \Lambda_{\alpha} \rangle_N} - \langle \Lambda_{\beta} \rangle_N \\ \nonumber
&=& \frac{\mathrm{cov}_N(\Lambda_0, \Lambda_{\beta}) + \sum_{\alpha}\mathrm{cov}_N(\Lambda_{\alpha},\Lambda_{\beta})K_{\alpha}(x_{N+1})}{\langle \Lambda_0 \rangle_N + \sum_{\alpha}K_{\alpha}(x_{N+1})\langle \Lambda_{\alpha} \rangle_N}
\end{eqnarray}
\end{widetext}
where:
\begin{equation}
\mathrm{cov}_N (\Lambda_A, \Lambda_B) \equiv \langle \Lambda_A\Lambda_B \rangle_N - \langle \Lambda_A \rangle_N \langle \Lambda_B \rangle_N 
\end{equation}
is the covariance. The updated mean may be written in terms of the original mean as:
\begin{widetext}
\begin{equation}\label{updated_mean}
\langle \Lambda_{\beta} \rangle_{N+1} = \langle \Lambda_{\beta} \rangle_{N} + \frac{\mathrm{cov}_N(\Lambda_0, \Lambda_{\beta}) + \sum_{\alpha}K_{\alpha}(x_{N+1})\mathrm{cov}_N(\Lambda_{\alpha},\Lambda_{\beta})}{\langle \Lambda_0 \rangle_N + \sum_{\alpha}K_{\alpha}(x_{N+1})\langle \Lambda_{\alpha} \rangle_N}.
\end{equation}
\end{widetext}
It is interesting to note here that the addition of a highly significant candidate event of a certain astrophysical source category could increase the mean values of the count for that category by more than unity. Indeed, if $K_{\beta}(x_{N+1}) \gg \langle \Lambda_0 \rangle_N$, and $K_{\beta}(x_{N+1}) \gg K_{\alpha \neq \beta} (x_{N+1})$, as is the case for certain highly significant events of category $\beta$, then  Eq.~\ref{mean_diff} is approximated as:
\begin{equation}
\langle \Lambda_{\beta} \rangle_{N+1} - \langle \Lambda_{\beta} \rangle_{N}  \approx \frac{\mathrm{var}_N(\Lambda_{\beta})}{\langle \Lambda_{\beta} \rangle}
\end{equation}
where $\mathrm{var}_N(\Lambda_{\beta}) \equiv \mathrm{cov}_N(\Lambda_{\beta}, \Lambda_{\beta})$. If the variance exceeds the mean, then the addition of that highly significant event would cause an increase in the mean of more than one. This is not so surprising, though perhaps counter intuitive at first. In effect, the addition of a highly significant event of a certain source-category informs the posterior that the rate of events of that source-category is higher than was inferred from the previously available set of triggers.

\section{Recursive Counts Posterior}

Given distinct chunks of GW data over which the spacetime volume sensitivity is assumed to be constant, we can construct FGMC counts posteriors for each chunk separately. However, comparing candidate events directly between chunks is not meaningful, because in general the spacetime volume sensitivity will differ from chunk to chunk. Therefore, in order to construct a posterior with all the events from all the chunks, we require some form of weighting involving the spacetime volume sensitivities of the chunks. 

The key idea here is that while the Poisson expected counts (the $\Lambda$s), and the spacetime volume sensitivities (the $\langle VT \rangle$s), will change from chunk to chunk, what is assumed to remain constant between chunks is the astrophysical rate of compact binary mergers that we seek to determine. 

In the following, we put superscripts to various quantities to identify the chunk of data they correspond to, and the subscripts label the source-category type, as earlier.  

We define the total spacetime-volume sensitivity, across all chunks $c$, for an astrophysical source-category $\alpha$, as:
\begin{equation}
\langle VT \rangle^{tot}_{\alpha} = \sum_{c} \langle VT \rangle_{\alpha}^{c}
\end{equation}
and the total Poisson expected counts for category $\alpha$ as:
\begin{equation}
\Lambda^{tot}_{\alpha} = \sum_{c} \Lambda^c_{\alpha}
\end{equation}
Working with the ``astrophysical-rate is time-independent'' assumption, the following must be true:
\begin{equation}
R_{\alpha} = \frac{\Lambda^{tot}_{\alpha}}{\langle VT \rangle^{tot}_{\alpha}} = \frac{\Lambda^c_{\alpha}}{\langle VT \rangle^c_{\alpha}}, \forall c
\end{equation}
where $R_{\alpha}$ is the astrophysical rate for category $\alpha$.

An important point to make here is that the background counts $\Lambda^c_0$ across chunks are not connected in any way: there is no unchanging $R_0$ corresponding to these counts. We can, however, resort to the delta-function approximation, and fix the terrestrial counts, for each chunk $c$, to the total number of candidate events for that chunk: $\Lambda^c_0 \rightarrow N^c$.

One can now combine the likelihoods from different chunks to construct a single posterior on the total counts $\Lambda^{tot}_{\alpha}$ of each astrophysical source category by making the following change of variables:
\begin{equation}
\Lambda^c_{\alpha} = \frac{\langle VT \rangle^c_{\alpha}}{\langle VT \rangle^{tot}_{\alpha}}\Lambda^{tot}_{\alpha}
\end{equation}
We can now write the combined counts-posterior as:
\begin{widetext}
\begin{equation}\label{combined_counts_posterior}
p(\Lambda_0, \vec{\Lambda}^{tot} | \vec{x}) = p(\vec{\Lambda}^{tot})\left\lbrace\prod_{c}\prod_{j=1}^{N^c}[N^c +  \vec{\Lambda}^{tot}\cdot \vec{K}^c(x^c_j)]\right\rbrace e^{-\vec{\Lambda}^{tot}\cdot \vec{u}}
\end{equation}
\end{widetext}
where:
\begin{equation}
\vec{K}^c = \left\lbrace K_{\alpha} \times  \frac{\langle VT \rangle^c_{\alpha}}{\langle VT \rangle^{tot}_{\alpha}} \right\rbrace, \alpha = 1,2,3,...,Q
\end{equation}
and:
\begin{equation}
\vec{\Lambda}^{tot} = \left\lbrace \Lambda^{tot}_{\alpha} \right\rbrace, \alpha = 1,2,3,...,Q
\end{equation}
with Q as the number of astrophysical source categories. 

It is worth noting here that the Gstlal pipeline incorporates spacetime volume sensitivity into its ranking statistic, and therefore the need to employ a method to weight candidate events by the sensitive spacetime volume sensitivity, as is done here, is redundant for the Gstlal pipeline. This however may not be true for all gravitational wave detection pipelines.

\section{Connecting terrestrial and astrophysical probabilities with mean values of the Poisson counts}\label{sec:appD}

Suppose the prior on the multi-component FGMC posterior (cf. Eq.~\eqref{counts_posterior}) has the form: $p(\Lambda_0, \vec{\Lambda}_1) = p(\Lambda_0,\Lambda_{\alpha}\notin\vec{\Lambda}_1) \Lambda_{\alpha}^a$. Taking the derivative of $\Lambda_{\alpha}p(\Lambda_0, \vec{\Lambda}_1~|~\vec{x})$ with respect to $\Lambda_{\alpha}$ yields:
\begin{equation}
\frac{d}{d\Lambda_{\alpha}}\left(\Lambda_{\alpha}p(\Lambda_0, \vec{\Lambda}_1~|~\vec{x})\right) = \left(a + 1\right)p(\Lambda_0, \vec{\Lambda}_1~|~\vec{x}) - \Lambda_{\alpha}p(\Lambda_0, \vec{\Lambda}_1~|~\vec{x}) + p(\Lambda_0, \vec{\Lambda}_1~|~\vec{x})\sum_{j=1}^{N}\frac{\Lambda_{\alpha}K_{\alpha}(x_j)}{\Lambda_0 + \vec{K}_{\alpha}(x_j)\cdot\vec{\Lambda}_{1}}
\end{equation}
The derivative was chosen so that
its antiderivative vanishes at $\Lambda_{\alpha} = 0$ and $\infty$ when marginalizing.

If the multi-component posterior is appropriately normalized, then, marginalizing out both sides of the above equation with respect to $\Lambda_0$ and $\vec{\Lambda}_1$, yields:
\begin{equation}
0 = a + 1 - \langle \Lambda_{\alpha} \rangle + \sum_{j=1}^{N} P_{\alpha}(x_j~|~\vec{x})
\end{equation}
which, upon rearranging, gives Eq~\eqref{mean_pastro}.

\end{document}